\documentclass[aps, amsmath, amssymb, reprint,superscriptaddress, showkeys]{revtex4-2}

\usepackage{dcolumn}
\usepackage{bm}
\usepackage{natbib}
\usepackage{color}
\usepackage{booktabs}  
\usepackage{siunitx}
\usepackage[final]{graphicx}
\usepackage{epstopdf}

\begin{document}

\preprint{APS/123-QED}

\title{\textbf{Quantum heat current in Terahertz-driven phonon systems} 
}%

\author{Yulong Qiao}
 
\author{R. Matthias Geilhufe}%
 
\affiliation{Department of Physics, Chalmers University of Technology, SE-41296, Gothenburg, Sweden
}%

\date{\today}

\begin{abstract}
The advent of high-intensity ultrafast laser pulses has opened new opportunities for controlling and designing quantum materials. In particular, terahertz (THz) pulses can resonantly drive optical phonon modes, enabling dynamic manipulation of lattice degrees of freedom. In this work, we investigate the ultrafast quantum thermodynamics of optical phonon mode driven by a THz pulse by treating the phonon as an open quantum system coupled to a thermal environment within a Caldeira–Leggett–type framework. We derive the quantum heat current between the phonon and the bath and analyze its behavior under realistic pulse protocols. Our results demonstrate that ultrafast laser driving can reveal and even induce significant deviations from the commonly adopted Markovian approximation, thereby providing a pathway to probe and control non-Markovian dissipation in driven solid-state systems.
\end{abstract}

\maketitle


\section{\label{Intro}Introduction}
Light–matter interactions provide a powerful means to manipulate and probe the internal states of quantum materials \cite{de2021}. In particular, terahertz (THz) laser pulses on the picosecond (ps) time scale have revolutionized the light-mediated control of lattice vibrations by resonantly driving selected phonon modes \cite{kampfrath2013}. A prototypical example is provided by the ferroelectric soft modes in perovskite oxides such as $\text{SrTiO}_3$ and $\text{KTaO}_3$, which have a high polarizability and a temperature-dependent resonance frequency in the THz range \cite{muller1979,han2007dielectric,ichikawa2005}. Recent advances in this field include the discovery of light-induced non-equilibrium metastable phases such as transient ferroelectricity~\cite{nova2019, cheng2023}, and emergent states such as dynamical multiferroicity~\cite{juraschek2017, basini2024}. Furthermore, the development of intense and ultrafast THz lasers enables large-amplitude phonon excitations, which in turn facilitate light-induced displacive phase transitions and complex energy transfer processes mediated by anharmonic phonon–phonon interactions \cite{mankowsky2017,kozina2019,disa2021}.

Theoretical descriptions of phonon dynamics driven by THz laser pulses have been largely established within classical frameworks \cite{shen1965,subedi2014,juraschek2017,radaelli2018,cheng2023}. Such models successfully reproduce phonon behavior at a phenomenological level. More recently, classical stochastic dynamics has been applied to evaluate thermodynamic quantities such as heat and entropy production in phonon systems by modeling microscopic fluctuations as stochastic forces~\cite{caprini2024}. However, at low temperatures or in materials containing light atoms, classical dynamics becomes insufficient, and quantum nuclear effects must be taken into account~\cite{rossi2021,caruso2023}. A well-known example is that the ferroelectric phase transition of $\text{SrTiO}_3$ predicted by classical theory is suppressed by quantum fluctuations at low temperature \cite{muller1979}. In this regime, the phonon mode has to be treated quantum mechanically and zero-point motion must be taken into account \cite{barrett1952,shin2021}. Moreover, an additional quantum contribution arises from the thermal bath to which the phonon couples: as temperature decreases, classical thermal noise is progressively suppressed, while quantum noise remains and can even dominate the dynamics. To gain a comprehensive picture, we construct a microscopic quantum phonon–bath model based on the paradigmatic Caldeira–Leggett framework and use it to investigate the non-equilibrium quantum thermodynamics of the optical phonon, in particular how quantum heat is transferred between the phonon and the bath under ultrafast THz driving. 


 On the other hand, the relevant phonon mode can strongly couple to other slow degrees of freedom with long correlation times, such as the emerging central peak with slow relaxation that appears when approaching the critical temperature of a soft mode associated with a structural phase transition \cite{shapiro1972,shirane1993,maity2025}. From the perspective of open quantum systems, such couplings render the effective bath non-Markovian, i.e., endowed with memory effects, so that describing the dissipation of the phonon with a constant decay rate inevitably misses important physics. In particular, we show that these non-Markovian features, both in the phonon dynamics and in the associated heat current, can be revealed and characterized through the response to an ultrafast laser pulse.

The remainder of this paper is organized as follows. In Sec.~\ref{sec:model}, we introduce the driven phonon mode coupled to a structured bath within the Caldeira–Leggett framework and introduce the quantum Langevin equation in the Heisenberg picture. In Sec.~\ref{sec:heat}, we formulate a general expression for the quantum heat current and specialize it to the driven phonon system. In Sec.~\ref{sec:numerical}, we specify the Lorentzian bath spectrum, and analyze the resulting non-Markovian phonon dynamics and quantum heat current under ultrafast driving. Conclusions and an outlook are given in Sec.~\ref{sec:conclusion}.

\section{Driven quantum phonon coupled to a quantum bath}\label{sec:model}

In order to describe the interaction between the phonon system and the laser pulse, as well as the dissipation of phonons, we model the whole setup by the Hamiltonian
\begin{equation}
    H = H_{\mathrm{ph}}(t) + H_{\mathrm{pB}}.
\end{equation}
The time-dependent Hamiltonian $H_{\mathrm{ph}}(t)$ describes the driven phonon mode within the harmonic approximation,
\begin{align}\label{eq:Hph}
    H_{\mathrm{ph}}(t) 
       &= \tfrac{\hat{P}^2}{2} 
       + \tfrac{\omega_{0}^2}{2} \hat{Q}^2 - F(t)\hat{Q},
\end{align}
where the first two terms correspond to the free Hamiltonian of a harmonic phonon mode with frequency $\omega_{0}$ at the $\Gamma$ point of the Brillouin zone (crystal momentum $\boldsymbol{q}=0$). Here $\hat{Q}$ and $\hat{P}$ denote the mass-weighted normal-mode coordinate and its conjugate momentum for the selected optical phonon at the $\Gamma$ point. They obey the canonical commutation relation 
\begin{align}
    [\hat{Q},\hat{P}] &= i\hbar,
\end{align}
with $\hbar$ denoting the reduced Planck constant. We note that anharmonic contributions such as a $\hat{Q}^4$ term can, in principle, be incorporated perturbatively. In the present work, however, we focus on the purely harmonic limit, which admits an elegant and fully analytical treatment.

\begin{figure}[htbp]
  \centering
  \includegraphics[width=1.0\linewidth]{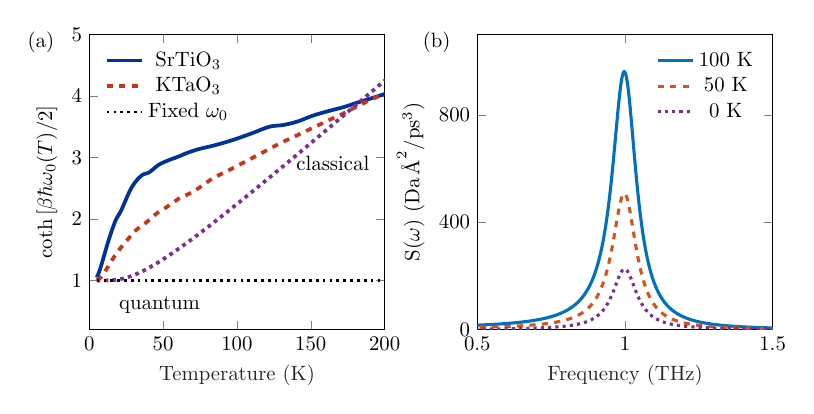}
  \caption{(a) Quantum criterion $\coth\left[\hbar\omega_0(T)/(2k_{\mathrm{B}}T)\right]$ for the ferroelectric soft mode of $\text{SrTiO}_3$ (blue solid), $\text{KTaO}_3$ (red dashed) and a fixed frequency $\omega_0=2.0~\text{THz}$ (purple dotted). The data for the temperature-dependent soft-mode frequency is taken from Ref.~\cite{vogt1995}. (b) Power spectral density $S(\omega)$ of the quantum noise for different temperatures. The bath spectral density $J(\omega)$ is taken to be Lorentzian as defined in Eq.~\eqref{eq:Jw}, The parameters are set to be $\Omega = 1.0~\text{THz}$, $\Gamma = 0.1\,\Omega$ and $g = 0.3\,\Omega^2$.}
  \label{fig:quantum_sto_kto}
\end{figure}

The quantum nature of the phonon mode is directly reflected in the fluctuations of the normal-mode coordinate, $\langle \hat{Q}^2\rangle$. For a harmonic oscillator in a Gibbs state at temperature $T$, one has
\[
    \langle \hat{Q}^2\rangle
    = \frac{\hbar}{2\omega}\coth\!\left(\frac{\beta\hbar\omega}{2}\right),
\]
where $\beta = 1/(k_{\mathrm{B}}T)$ and $k_{\mathrm{B}}$ is the Boltzmann constant. As the temperature approaches zero, the hyperbolic cotangent converges to unity and the fluctuation saturates at
$\langle \hat{Q}^2\rangle = \hbar/(2\omega)$, which corresponds to the zero-point motion of the mode. In Fig.~\ref{fig:quantum_sto_kto}a, we take the ferroelectric soft modes of $\text{SrTiO}_3$ and $\text{KTaO}_3$ as examples and use their temperature-dependent frequencies $\omega_0(T)$ to evaluate the factor $\coth[\beta\hbar\omega_0(T)/2]$. For comparison, a fixed frequency $\omega_0 = 2.0~\text{THz}$ is also considered as a reference. It is evident that at low temperature this quantity approaches unity in all three cases rather than vanishing, indicating the dominance of zero-point motion. The lower values obtained for KTaO$_3$ at a given temperature reflect its larger soft-mode frequency compared to SrTiO$_3$ \cite{vogt1995}, while the fixed-frequency case approaches unity most rapidly, consistent with its even higher characteristic energy scale. In the high-temperature limit $k_{\mathrm{B}}T \gg \hbar\omega_0$, by contrast, $\coth(\beta\hbar\omega_0/2)\approx 2k_{\mathrm{B}}T/(\hbar\omega_0)$, so that the average potential energy reduces to the classical result $k_{\mathrm{B}}T/2$.

The last term in Eq.~\eqref{eq:Hph} describes the coupling of the optical phonon to a time-dependent Gaussian laser force,
\begin{align}
    F(t) = ZE_0 \exp\left[-\frac{(t-\tau_0)^2}{2\tau^2}\right]\cos(\omega_0 t),
\end{align}
where $E_0$ is electric field strength in the medium and $Z$ is the mode effective charge \cite{gonze1997}, while $\tau_0$ denotes the time at which the pulse reaches its maximum, and $\tau$ characterizes the duration of the Gaussian envelope. The wave vector of the THz laser is vanishingly small compared to the crystal momentum scale, such that the driving couples only to phonons at the $\Gamma$ point.

In reality, the optical phonon is inevitably coupled to other microscopic degrees of freedom, such as lattice vibrations and electrons, which leads to dissipation. These additional degrees of freedom can be coarse-grained into an effective thermal bath. The second part of the total Hamiltonian, $H_{\mathrm{pB}}$, incorporates the influence of this bath on the phonon dynamics at low temperature. To this end, we employ the paradigmatic Caldeira--Leggett model \cite{caldeira1983,hu1992} to describe the interaction between the optical phonon and the bath,  
\begin{align}
    H_{\mathrm{pB}}
    = \sum_{k}\left[
        \frac{\hat{p}_k^2}{2}
        + \frac{1}{2}\omega_k^2\left(\hat{q}_k - \frac{g_k}{\omega_k^2}\hat{Q}\right)^2
    \right],
\end{align}
where the bath consists of a set of harmonic oscillators with frequencies $\omega_k$. Here $\hat{q}_k$ and $\hat{p}_k$ denote the mass-weighted coordinate and its conjugate momentum for the $k$th bath mode, respectively, and they obey the canonical commutation relations $[\hat{q}_k,\hat{p}_{k'}]=i\hbar\,\delta_{kk'}$. The parameter $g_k$ quantifies the coupling strength between the system coordinate $\hat{Q}$ and the $k$th bath mode. We have assumed a linear, harmonic coupling of the form $g_k \hat{q}_k \hat{Q}$ together with the standard counter-term $\frac{g_k^2}{2\omega_k^2}\hat{Q}^2$, which compensates the frequency shift of the system induced by the bath and keeps $\omega_0$ as the physical phonon frequency.

In the Heisenberg picture, the equation of motion for the normal-mode coordinate $\hat{Q}(t)$ of the optical phonon can be written as \cite{ford1988,yu1994,sun1995exact,zhang2024}
\begin{align}\label{eq:QIR2}
    \mathcal{L}\hat{Q}(t) = F(t) + \hat{\xi}(t),
\end{align}
where the linear integro-differential operator $\mathcal{L}$ is defined by
\begin{align}
    \mathcal{L}\hat{Q}(t)
    &\equiv \ddot{\hat{Q}}(t) + \omega_{0}^2 \hat{Q}(t) \notag \\
    &\quad + \int_{t_0}^{t} \mathrm{d}s\, \gamma(t-s)\,\dot{\hat{Q}}(s)
      + \gamma(t-t_0)\,\hat{Q}(t_0).
    \label{eq:QIR3} 
\end{align}
Equation~\eqref{eq:QIR2} is a generalized quantum Langevin equation, in which the friction kernel $\gamma(t)$ is given by
\begin{align}\label{eq:gamma_t}
    \gamma(t)
    = \sum_{k}\frac{g_k^2}{\omega_k^2}\cos(\omega_k t)
    = \int_{0}^{\infty}\mathrm{d}\omega\, \frac{J(\omega)}{\omega}\cos(\omega t),
\end{align}
with the spectral density
\begin{align}
    J(\omega)
    = \sum_k \frac{g_k^2}{\omega_k}\,\delta(\omega-\omega_k),
\end{align}
whose specific form is determined by the interaction information between the phonon and the bath.
The operator $\hat{\xi}(t)$ represents the quantum noise originating from the bath,
\begin{align}
    \hat{\xi}(t)
    = \sum_{k}\left[
        g_k \hat{q}_k(0)\cos(\omega_k t)
        + \frac{g_k}{\omega_k}\hat{p}_k(0)\sin(\omega_k t)
    \right],
\end{align}
whose ensemble average vanishes, $\langle \hat{\xi}(t) \rangle = 0$, provided that the bath is initially in thermal equilibrium, e.g., in a Gibbs state. We note that if the initial time $t_0$ is taken to the remote past, the last term in Eq.~\eqref{eq:QIR3}, $\gamma(t-t_0)\hat{Q}(t_0)$, can be neglected.

The influence of the noise operator on the system dynamics is encoded in its statistical properties. Its correlation function $\eta(t)$ is defined via \cite{gardiner2004},
\begin{align}\label{eq:anticomm}
    \eta(t-t')
    &= \big\langle\{\hat{\xi}(t),\hat{\xi}(t')\}\big\rangle \nonumber\\
    &= \hbar \int_0^{\infty}\mathrm{d}\omega\, 
       J(\omega)\coth\!\left(\frac{\beta\hbar\omega}{2}\right)
       \cos\big[\omega(t-t')\big],
\end{align}
where $\{\cdots\}$ denotes the anti-commutator. In Eq.~\eqref{eq:anticomm} we have assumed that the thermal bath is initially prepared in a Gibbs state at temperature $T$. The hyperbolic cotangent factor $\coth(\beta\hbar\omega/2)$ encompasses both thermal and quantum (zero-point) fluctuations of the bath modes.

The power spectral density of the quantum noise is given by the Fourier transform of the correlation function,
\begin{align}\label{eq:Sw}
    S(\omega)
    &= \int_{-\infty}^{\infty}\mathrm{d}t\, \eta(t)\, e^{-i\omega t} \nonumber\\
    &= \pi\hbar\, J(|\omega|)\coth\!\left(\frac{\beta\hbar|\omega|}{2}\right).
\end{align}
By comparing Eq.~\eqref{eq:Sw} with the definition of the friction kernel in Eq.~\eqref{eq:gamma_t}, we obtain the quantum fluctuation--dissipation theorem
\begin{align}\label{eq:FD_quantum}
    S(\omega)
    = \hbar\omega\,\coth\!\left(\frac{\beta\hbar\omega}{2}\right)\tilde{\gamma}(|\omega|),
\end{align}
where $\tilde{\gamma}(\omega)$ denotes the Fourier transform of $\gamma(t)$. In the high-temperature regime, i.e., when $k_{\mathrm{B}}T$ is much higher than any relevant energy scales of the bath modes, the hyperbolic cotangent can be expanded and one recovers
\begin{align}\label{eq:classical_FD}
    S(\omega) \simeq 2k_{\mathrm{B}}T\,\tilde{\gamma}(|\omega|),
\end{align}
which is the classical form of the fluctuation--dissipation theorem. In Fig.~\ref{fig:quantum_sto_kto}b, we present $S(\omega)$ as the function of the frequency with different temperatures. The spectral density $J(\omega)$ is set to be Lorentzian which will be explicitly introduced later. It is clear that even though the temperature is zero, the quantum noise will not vanish which is in contrast to the classical prediction.


Using a Laplace transform, the formal solution of Eq.~\eqref{eq:QIR3} for the normal-mode coordinate can be written as
\begin{align}\label{eq:QIRO}
    \hat{Q}(t)
    &= \dot{G}(t)\,\hat{Q}(t_0)
     + G(t)\,\dot{\hat{Q}}(t_0) \notag\\ 
    &\quad + \int_{t_0}^t \mathrm{d}s\, G(t-s)\big[\hat{\xi}(s)+F(s)\big],
\end{align}
where $G(t)$ is the retarded Green's function (response function) determined by
\begin{align}
    \ddot{G}(t)
    + \omega_{0}^2 G(t)
    + \int_{t_0}^t\mathrm{d}s\,\gamma(t-s)\,\dot{G}(s) = 0,
\end{align}
with initial conditions $G(t_0)=0$ and $\dot{G}(t_0)=1$.

\section{Quantum heat current in a driven phonon system} \label{sec:heat}

The laser force drives the phonon system out of equilibrium and induces heat exchange between the system and the bath. In contrast to classical systems, in the quantum regime the heat cannot, in general, be uniquely separated from other contributions to the total energy because of quantum correlations between the system and the bath \cite{talkner2020}. However, when the interaction between the thermal bath and the phonon system is sufficiently weak so that the energy stored in the interaction Hamiltonian can be neglected, the heat dissipated into the bath can be well approximated by the heat absorbed by the system, as discussed in detail in Appendix~\ref{appendix:heat}. 

To avoid ambiguity, in the following we define the quantum heat $\langle\mathcal{Q}\rangle$ as the energy flowing from the thermal bath to the phonon system. Here we use the calligraphic symbol $\mathcal{Q}$ for the heat in order to distinguish it from the phonon normal-mode coordinate $\hat{Q}$.  The calculation of this heat does not rely on the weak-coupling approximation and can be derived exactly. A general derivation presented in Appendix~\ref{appendix:heat}, shows that the quantum heat can be written as
\begin{align}
    \langle\mathcal{Q}(\tau)\rangle
    = \int_{t_0}^{\tau}\mathrm{d}t\, J(t),
\end{align}
where the quantum heat current $J(t)$ is given by
\begin{align}\label{eq:heat_current}
    J(t)
    = \frac{\mathrm{d}}{\mathrm{d}t}\big\langle\tilde{H}_{\mathrm{ph}}(t)\big\rangle
      - \Big\langle\widetilde{\frac{\partial H_{\mathrm{ph}}(t)}{\partial t}}\Big\rangle.
\end{align}
Here we have introduced the time-dependent operators in the Heisenberg picture,
\begin{align}
    \tilde{H}_{\mathrm{ph}}(t) &= U^\dagger(t)\,H_{\mathrm{ph}}(t)\,U(t),\\[1mm]
    \widetilde{\frac{\partial H_{\mathrm{ph}}(t)}{\partial t}} 
    &= U^\dagger(t)\,\frac{\partial H_{\mathrm{ph}}(t)}{\partial t}\,U(t),
\end{align}
with $U(t)$ denoting the unitary time-evolution operator of the full system (phonon plus bath). We note that Eq.~\eqref{eq:heat_current} is derived for a general time-dependent Hamiltonian, as detailed in Appendix~\ref{appendix:heat}, and is not restricted to the specific form of Eq.~\eqref{eq:Hph}. 

Applying Eq.~\eqref{eq:heat_current} to the phonon Hamiltonian in Eq.~\eqref{eq:Hph}, we obtain
\begin{align}\label{eq:J_t}
    J(t)
    = -\big\langle F(t)\,\dot{\hat{Q}}(t)\big\rangle
      + \frac{1}{2}\Big\langle\Big\{\ddot{\hat{Q}}(t)+\omega_{0}^2 \hat{Q}(t),\dot{\hat{Q}}(t)\Big\}\Big\rangle.
\end{align}
The first term on the right-hand side of Eq.~\eqref{eq:J_t} corresponds to the rate at which work is done by the laser force on the full system, while the second term is the rate of change of the internal energy of the phonon subsystem. 

By substituting Eqs.~\eqref{eq:QIR2} and \eqref{eq:QIRO} into Eq.~\eqref{eq:J_t}, and assuming that $t_0$ is set to be the remote past and that the system and the bath are initially uncorrelated, the heat current can be recast in the form
\begin{align}\label{eq:J_t2}
    J(t)
    &= -\frac{1}{2}\int_{t_0}^t\mathrm{d}s\,\gamma(t-s)\,
        \big\langle\{\dot{\hat{Q}}(t),\dot{\hat{Q}}(s)\}\big\rangle \nonumber\\
    &\quad + \frac{1}{2}\int_{t_0}^t\mathrm{d}s\, \dot{G}(t-s)\,\eta(t-s),
\end{align}
where the first term on the right-hand side can be naturally interpreted as the dissipative contribution, governed by the friction kernel $\gamma(t)$, while the second term arises from bath fluctuations encoded in the noise correlation function $\eta(t)$. Equation~\eqref{eq:J_t2} represents a general expression for the heat current in a non-Markovian quantum dynamics. The non-Markovian character is reflected in the dependence on the history of the motion through the memory kernel $\gamma(t-s)$ and the Green's function $G(t-s)$, while the quantum nature of the problem is reflected in the operator structure and the use of anti-commutators. Technically, the explicit expression for the velocity of the phonon normal-mode coordinate $\dot{\hat{Q}}(t)$ is obtained by taking the time derivative of Eq.~\eqref{eq:QIRO}. Consequently, the anti-commutator $\{\dot{\hat{Q}}(t),\dot{\hat{Q}}(s)\}$ contains three types of contributions: (i) terms depending on the initial operators $\hat{Q}(t_0)$ and $\dot{\hat{Q}}(t_0)$, which contain the quantum fluctuation of the phonon mode, (ii) a term involving the laser force $F(t)F(s)$, and (iii) a term containing the noise correlation function $\eta(t-s)$.

In the Markovian limit, the friction kernel becomes local in time, $\gamma(t-s) = 2\gamma_0\,\delta(t-s)$, where $\delta(t)$ is the Dirac delta function and $\gamma_0$ is a constant decay rate. In addition, in the high-temperature limit the noise correlation reduces to $\eta(t-s) = 4\gamma_0 k_{\mathrm{B}}T\,\delta(t-s)$ according to Eq.~\eqref{eq:classical_FD}. Under these assumptions, Eq.~\eqref{eq:J_t2} simplifies to
\begin{align}\label{eq:Jt_Markovian}
    J(t) = -\gamma_0\big\langle\dot{\hat{Q}}^2(t)\big\rangle + \gamma_0 k_{\mathrm{B}}T,
\end{align}
where the first term depends only on the instantaneous velocity $\dot{\hat{Q}}(t)$, and the second term is a constant determined solely by the temperature and the decay rate. From Eq.~\eqref{eq:Jt_Markovian} we infer that $J(t)=0$ only when the system has relaxed to thermal equilibrium, where $\langle \dot{\hat{Q}}^2(t)\rangle = k_{\mathrm{B}}T$. Thus, the heat current provides a direct measure of how far the system is driven out of equilibrium.

\begin{figure}[htbp]
  \centering
  \includegraphics[width=1.0\linewidth]{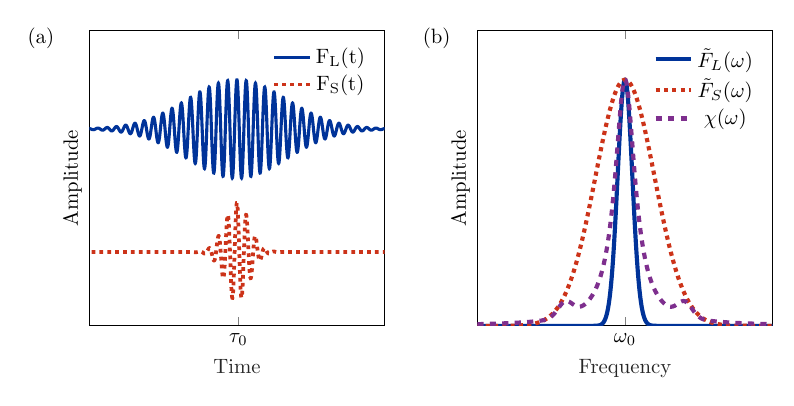}
  \caption{(a) Time profiles of a long Gaussian laser pulse $F_L(t)$ (blue solid) and a short Gaussian laser pulse $F_S(t)$ (red dotted), both centered at $t_0$. (b) Schematic illustration of the corresponding frequency distributions: the spectra of the long pulse $\tilde{F}_L(\omega)$ (blue solid), the short pulse $\tilde{F}_S(\omega)$ (red dotted) and the system’s response function $\chi(\omega)$ (purple dashed). 
}
  \label{fig:laser_width}
\end{figure}
\section{Numerical analysis of non-Markovian quantum heat current} \label{sec:numerical}

To specify the properties of the bath, we adopt a Lorentzian spectral density of the Brownian-oscillator type,
\begin{align}\label{eq:Jw}
    J(\omega)
    = \frac{1}{\pi}\,
      \frac{g^2 \Gamma\,\omega}
           {\big(\omega^2-\Omega^2\big)^{2}
            + \Gamma^2 \omega^2},
\end{align}
where $g$ is the coupling strength between the system and the bath, and $\Gamma$ determines the width of the distribution centered at frequency $\Omega$. This structured spectrum captures the physical situation in which the phonon system is strongly coupled to a specific bath mode that can itself be modeled as a damped harmonic oscillator \cite{barra2021}. In time domain, $1/\Gamma$ determines the relaxation time of both the frictional kernel $\gamma(t)$ and the correlation function $\eta(t)$ \cite{anders2022}. Thus, a smaller $\Gamma$ corresponds to a longer bath correlation time and therefore enhances non-Markovian effects. We also note that the spectral density can take different forms, depending on the details of the interaction between the phonon and other ionic degrees of freedom or electrons.

In this setting, the dynamics of the full system are governed by three characteristic time scales: the intrinsic time scale of the phonon mode, $1/\omega_0$, the bath correlation time, $1/\Gamma$, and the pulse duration of the laser, $\tau$. The relative magnitude of these time scales can significantly influence the observed dynamics. To make their impact more transparent, it is convenient to work in Fourier space. The solution of Eq.~\eqref{eq:QIR3} can then be written formally as
\begin{align}
    \tilde{\hat{Q}}(\omega)
    = \chi(\omega)\big[\tilde{F}(\omega)+\tilde{\hat{\xi}}(\omega)\big],
\end{align}
where $\tilde{F}(\omega)$ and $\tilde{\hat{\xi}}(\omega)$ are the Fourier transforms of $F(t)$ and $\hat{\xi}(t)$, respectively. The frequency response function $\chi(\omega)$ is given by
\begin{align}
    \chi(\omega)
    = \frac{1}{\omega_{0}^2-\omega^2
               + i\omega\tilde{\gamma}(\omega)},
\end{align}
where $\tilde{\gamma}(\omega)$ is the (single-sided) Fourier transform of the friction kernel $\gamma(t)$, which in turn is determined by $J(\omega)$ through Eq.~\eqref{eq:gamma_t}. This response function shows that the system responds most strongly in a frequency window around $\omega_0$, with a width set by the detailed structure of $\tilde{\gamma}(\omega)$. If $\tilde{\gamma}(\omega)$ exhibits a pronounced frequency dependence in the vicinity of $\omega_0$ which happens for a small $\Gamma$, the resulting dynamics becomes non-Markovian. In contrast, if $\tilde{\gamma}(\omega)$ is effectively constant over the relevant frequency range, corresponding to a large $\Gamma$ and a time-local friction kernel $\gamma(t)\propto\delta(t)$, the response function $\chi(\omega)$ reduces to a simple Lorentzian and the dynamics are Markovian.

Although Markovianity is an intrinsic property of the system--bath coupling, it can be probed and revealed by an external driving field. For a Gaussian laser pulse $F(t)$, whose Fourier transform is also Gaussian, a short duration $\tau$ implies that $\tilde{F}(\omega)$ extends over a wide frequency range due to the uncertainty relation \[\Delta\omega\,\Delta\tau\approx\text{const},\] approximately covering the interval $[\omega_{0}-2\pi/\tau,\;\omega_{0}+2\pi/\tau]$. In Fig.~\ref{fig:laser_width}a, we draw a long Gaussian pulse and a short Gaussian pulse. Their corresponding Fourier transformations are presented in Fig.~\ref{fig:laser_width}b, where the long pulse has a narrow frequency distribution while the short pulse is associated with a much broader spectrum. Furthermore, as illustrated in Fig.~\ref{fig:laser_width}b, if $\chi(\omega)$ deviates significantly from a Lorentzian within the frequency window of the short pulse $\tilde{F}_S(\omega)$ because of the nontrivial structure of $J(\omega)$, many frequency components of $\chi(\omega)$ are excited and the underlying non-Lorentzian spectrum can be resolved. In contrast, for the long pulse, $\tilde{F}_L(\omega)$ becomes sharply peaked around $\omega_0$, such that $\chi(\omega)\approx \chi(\omega_0)$ within the regime of $\tilde{F}_L(\omega)$ and the frequency dependence of $J(\omega)$ is no longer effectively probed.

\begin{figure}
  \centering
  \includegraphics[width=1.0\linewidth]{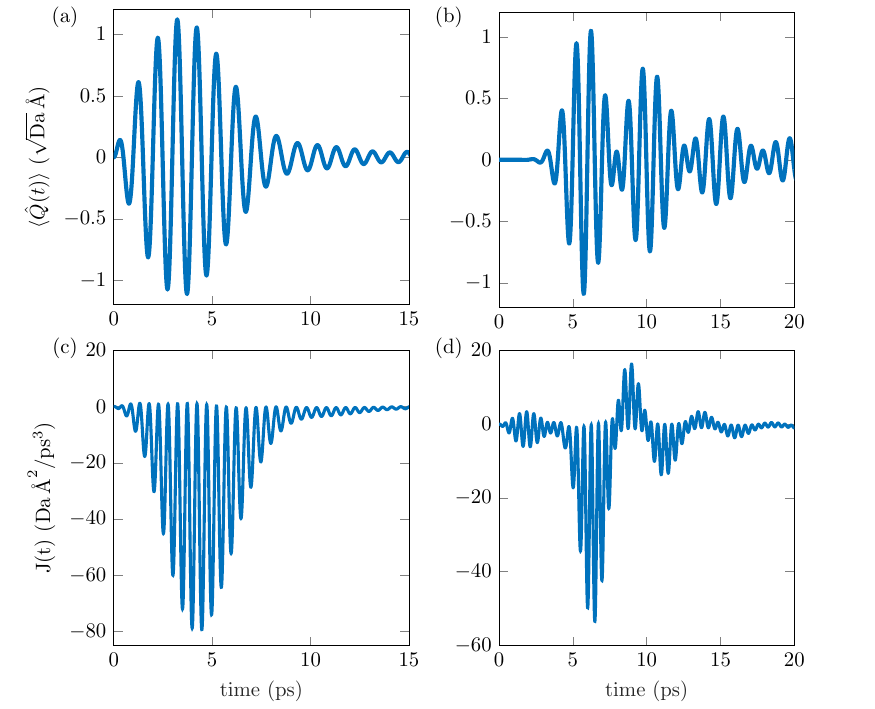}
  \caption{Influence of the pulse duration $\tau$ of the laser force on the dynamics of $\langle \hat{Q}(t)\rangle$ and on the quantum heat current $J(t)$. (a) Dynamics of the phonon displacement $\langle \hat{Q}(t)\rangle$ driven by a long pulse with $\tau = 5.0~\text{ps}$; (b) $\langle \hat{Q}(t)\rangle$ driven by a short pulse with $\tau = 1.0~\text{ps}$; (c) quantum heat current calculated from Eq.~\eqref{eq:J_t2} for the long pulse; and (d) the corresponding heat current for the short pulse. The fixed parameters of the spectral density $J(\omega)$ are $\Gamma = 0.1~\omega_{0}$ and $g = 0.3~\omega_{0}^2$.}
  \label{fig:Q_J}
\end{figure}

To test this idea, we analyze the influence of the pulse duration $\tau$ on the dynamics of $\langle \hat{Q}(t)\rangle$ and on the quantum heat current $J(t)$, as shown in Fig.~\ref{fig:Q_J}. For simplicity, we assume that the initial states of the phonon and the bath factorize and that both are prepared in a Gibbs state at the same temperature. The phonon frequency is chosen as $\omega_{0}/2\pi=\Omega/2\pi=1.0~\text{THz}$, and the temperature is set to $T=70~\text{K}$, which are consistent with experimental values for the ferroelectric soft mode in $\text{SrTiO}_3$ \cite{vogt1995}. The amplitude of the laser force is fixed to $ZE_0 = 10~\sqrt{\si{\dalton}}\;\si{\angstrom}/\si{\pico\second\squared}$ in atomic units, which is a moderate value compared to state-of-the-art experiments \cite{kozina2019}. The peak of the laser force appears at $\tau_0=5~\text{ps}$. For the Lorentzian spectral density $J(\omega)$ we take $\Gamma = 0.1~\omega_{0}$ and $g = 0.3~\omega_{0}^2$, which corresponds to a strongly structured, and thus strongly non-Markovian, spectrum.

For a long pulse with $\tau = 5.0~\text{ps}$, the amplitude of $\langle \hat{Q}(t)\rangle$ shown in Fig.~\ref{fig:Q_J}(a) decays monotonically after the pulse, resembling a typical Markovian relaxation even though the underlying $J(\omega)$ is non-Markovian. In this case, the narrow frequency distribution of the driving probes only a small region around $\omega_0$, where $\chi(\omega)$ is effectively Lorentzian. In stark contrast, for a short pulse with $\tau = 1.0~\text{ps}$, the broad frequency distribution of the driving leads to a distinctly non-Markovian dynamics. As shown in Fig.~\ref{fig:Q_J}(b), the amplitude of $\langle \hat{Q}(t)\rangle$ decays rapidly to nearly zero after the pulse has passed and then exhibits revivals reminiscent of quantum beats. This behavior demonstrates that the nontrivial structure of $J(\omega)$ can be resolved by an ultrafast pulse and is clearly imprinted in the phonon dynamics. 

The same crossover is also reflected in the quantum heat current. In Fig.~\ref{fig:Q_J}(c), the heat current $J(t)$ induced by the long pulse shows that energy flows from the phonon into the bath in an essentially unidirectional way, indicating that bath memory effects are negligible on the probed time scales. By contrast, in Fig.~\ref{fig:Q_J}(d) for the short-pulse case, $J(t)$ changes sign and energy can flow back from the bath to the phonon, which is a characteristic signature of non-Markovian dynamics and coincides with the revivals observed in Fig.~\ref{fig:Q_J}(b). As a simple and experimentally accessible dynamical indicator of non-Markovianity, we quantify the ratio between the energy that flows back from the bath to the phonon and the total exchanged energy after the pulse as
\begin{align}
    \mathcal{M}_J
    = \frac{\displaystyle\int_{t_p}^{\tau_{\max}}\!\mathrm{d}t\, \max\{0, J(t)\}}
           {\displaystyle\int_{t_p}^{\tau_{\max}}\!\mathrm{d}t\, |J(t)|},
\end{align}
where $t_p$ denotes the end of the driving pulse and $\tau_{\max}$ is the time at which the phonon has relaxed back to equilibrium. In the Markovian limit, the heat current is essentially unidirectional due to the friction term $-\gamma_0\langle \dot{\hat{Q}}^2(t)\rangle$ in Eq.~\eqref{eq:Jt_Markovian}, and one finds $\mathcal{M}_J \simeq 0$. In contrast, a strongly non-Markovian bath leads to significantly positive contributions of $J(t)$ and hence to a finite value of $\mathcal{M}_J$. A similar indicator for the degree of non-Markovianity based on the negative entropy-production rate has been proposed in \cite{strasberg2019,Hartmann2025_inprep}.

Finally, we emphasize that the heat current calculated here genuinely involves quantum effects. Although in both classical and quantum descriptions the heat current $J(t)$ vanishes once the phonon has equilibrated with the bath, the underlying balance between the dissipative contribution and the fluctuation contribution is qualitatively different. In the classical case, the two contributions scale with temperature and therefore go to zero as $T \to 0$, as shown in Eq.~\eqref{eq:Jt_Markovian}. In the quantum case, by contrast, the quantum fluctuation–dissipation relation contains the factor $\coth(\beta\hbar\omega/2)$ encoded in Eq.~\eqref{eq:J_t2}, which approaches unity at low temperature. As a result, both contributions saturate at finite values set by zero-point motion, even though their sum leads to a vanishing net heat current in equilibrium.

\section{Conclusion}\label{sec:conclusion}

In this work, we have developed a microscopic quantum description of an optical phonon mode driven by an ultrafast THz laser pulse and coupled to a thermal environment modeled by a Caldeira–Leggett–type Hamiltonian, such that the phonon dynamics is governed by a generalized quantum Langevin equation. On this basis, we derived a general expression for the heat current in non-Markovian quantum dynamics and analyzed how a structured bath and ultrafast driving impact the energy exchange between the phonon and its environment. Our results demonstrate that ultrafast THz pulses provide a sensitive probe of non-Markovianity in the phonon dynamics.

For the outlook, we note that the non-Markovian dynamics of optical phonons discussed here can be accessed in pump--probe experiments. A particularly promising candidate is the ferroelectric soft mode which can be strongly coupled to other slow degrees of freedom at low temperature, even though the harmonic oscillator remains an idealization for soft mode in real materials. In such a setting, a THz Gaussian pulse can be used to resonantly excite the selected optical phonon mode while systematically shortening the pulse duration on the picosecond time scale. The ensuing lattice motion can be monitored by ultrafast x-ray diffraction with femtosecond time resolution \cite{kozina2019}. A characteristic signature of non-Markovianity in this context would be an anomalous re-acceleration or revival of the coherent lattice motion after the driving pulse has essentially vanished, indicating a transient backflow of energy from the bath to the phonon mode. Such experimentally observable non-Markovian behavior would provide direct access to detailed information about the phonon–bath interaction encoded in the bath spectral density.


\begin{acknowledgments}
We acknowledge support from the Swedish Research Council (VR starting Grant No. 2022-03350), the Olle Engkvist Foundation (Grant No. 229-0443), the Royal Physiographic Society in Lund (Horisont), the Knut and Alice Wallenberg Foundation (Grant No. 2023.0087), and Chalmers University of Technology, via the department of physics and the Areas of Advance Nano and Materials Science. 
\end{acknowledgments}

\bibliography{apssamp}

@article{nova2019,
  title={Metastable ferroelectricity in optically strained {$\mathrm{SrTiO_3}$}},
  author={Nova, TF and Disa, AS and Fechner, Michael and Cavalleri, Andrea},
  journal={Science},
  volume={364},
  number={6445},
  pages={1075--1079},
  year={2019},
  publisher={American Association for the Advancement of Science}
}

@article{cheng2023,
  title={Terahertz-driven local dipolar correlation in a quantum paraelectric},
  author={Cheng, Bing and Kramer, Patrick L and Shen, Zhi-Xun and Hoffmann, Matthias C},
  journal={Physical Review Letters},
  volume={130},
  number={12},
  pages={126902},
  year={2023},
  publisher={APS}
}

@article{juraschek2017,
  title={Dynamical multiferroicity},
  author={Juraschek, Dominik M and Fechner, Michael and Balatsky, Alexander V and Spaldin, Nicola A},
  journal={Physical Review Materials},
  volume={1},
  number={1},
  pages={014401},
  year={2017},
  publisher={APS}
}

@article{basini2024,
  title={Terahertz electric-field-driven dynamical multiferroicity in {$\mathrm{SrTiO_3}$}},
  author={Basini, Martina and Pancaldi, Matteo and Wehinger, Bj{\"o}rn and Udina, Mattia and Unikandanunni, Vivek and Tadano, Terumasa and Hoffmann, Matthias C and Balatsky, Alexander V and Bonetti, Stefano},
  journal={Nature},
  volume={628},
  number={8008},
  pages={534--539},
  year={2024},
  publisher={Nature Publishing Group UK London}
}

@article{mankowsky2017,
  title={Ultrafast reversal of the ferroelectric polarization},
  author={Mankowsky, Roman and von Hoegen, Alexander and F{\"o}rst, Michael and Cavalleri, Andrea},
  journal={Physical review letters},
  volume={118},
  number={19},
  pages={197601},
  year={2017},
  publisher={APS}
}

@article{kozina2019,
  title={Terahertz-driven phonon upconversion in {$\mathrm{SrTiO_3}$}},
  author={Kozina, Michael and Fechner, Michael and Marsik, Premysl and van Driel, Tim and Glownia, James M and Bernhard, Christian and Radovic, Milan and Zhu, Diling and Bonetti, Stefano and Staub, Urs and others},
  journal={Nature Physics},
  volume={15},
  number={4},
  pages={387--392},
  year={2019},
  publisher={Nature Publishing Group UK London}
}

@article{disa2021,
  title={Engineering crystal structures with light},
  author={Disa, Ankit S and Nova, Tobia F and Cavalleri, Andrea},
  journal={Nature Physics},
  volume={17},
  number={10},
  pages={1087--1092},
  year={2021},
  publisher={Nature Publishing Group UK London}
}

@article{caprini2024,
  title={Ultrafast entropy production in pump-probe experiments},
  author={Caprini, Lorenzo and L{\"o}wen, Hartmut and Geilhufe, R Matthias},
  journal={Nature Communications},
  volume={15},
  number={1},
  pages={94},
  year={2024},
  publisher={Nature Publishing Group UK London}
}

@article{caruso2023,
  title={Quantum theory of light-driven coherent lattice dynamics},
  author={Caruso, Fabio and Zacharias, Marios},
  journal={Physical Review B},
  volume={107},
  number={5},
  pages={054102},
  year={2023},
  publisher={APS}
}

@article{caldeira1983,
  title={Path integral approach to quantum Brownian motion},
  author={Caldeira, Amir O and Leggett, Anthony J},
  journal={Physica A: Statistical mechanics and its Applications},
  volume={121},
  number={3},
  pages={587--616},
  year={1983},
  publisher={Elsevier}
}

@article{hu1992,
  title={Quantum Brownian motion in a general environment: Exact master equation with nonlocal dissipation and colored noise},
  author={Hu, Bei Lok and Paz, Juan Pablo and Zhang, Yuhong},
  journal={Physical Review D},
  volume={45},
  number={8},
  pages={2843},
  year={1992},
  publisher={APS}
}

@article{ford1988,
  title={Quantum langevin equation},
  author={Ford, George W and Lewis, John T and O’Connell, RF943786},
  journal={Physical Review A},
  volume={37},
  number={11},
  pages={4419},
  year={1988},
  publisher={APS}
}

@article{yu1994,
  title={Evolution of the wave function in a dissipative system},
  author={Yu, Li Hua and Sun, Chang-Pu},
  journal={Physical Review A},
  volume={49},
  number={1},
  pages={592},
  year={1994},
  publisher={APS}
}

@article{sun1995exact,
  title={Exact dynamics of a quantum dissipative system in a constant external field},
  author={Sun, Chang-Pu and Yu, Li-Hua},
  journal={Physical Review A},
  volume={51},
  number={3},
  pages={1845},
  year={1995},
  publisher={APS}
}

@article{zhang2024,
  title={Statistics of quantum heat in the Caldeira-Leggett model},
  author={Zhang, Ze-Zhou and Tan, Qing-Shou and Wu, Wei},
  journal={Physical Review E},
  volume={109},
  number={6},
  pages={064134},
  year={2024},
  publisher={APS}
}

@book{gardiner2004,
  title={Quantum noise: a handbook of Markovian and non-Markovian quantum stochastic methods with applications to quantum optics},
  author={Gardiner, Crispin and Zoller, Peter},
  year={2004},
  publisher={Springer Science \& Business Media}
}

@article{vogt1995,
  title={Refined treatment of the model of linearly coupled anharmonic oscillators and its application to the temperature dependence of the zone-center soft-mode frequencies of {$\mathrm{KTaO_3}$} and {$\mathrm{SrTiO_3}$}},
  author={Vogt, H},
  journal={Physical Review B},
  volume={51},
  number={13},
  pages={8046},
  year={1995},
  publisher={APS}
}

@article{maity2025,
  title={Soft phonon and the central peak at the cubic-to-tetragonal phase transition in {$\mathrm{SrTiO_3}$}},
  author={Maity, Avishek and Habicht, Klaus and Merz, Michael and Said, Ayman H and Guguschev, Christo and Kojda, Danny and Ryll, Britta and Hoffmann, Jan-Ekkehard and Dittmar, Andrea and Keller, Thomas and others},
  journal={Physical Review B},
  volume={111},
  number={13},
  pages={134108},
  year={2025},
  publisher={APS}
}

@article{radaelli2018,
  title={Breaking symmetry with light: Ultrafast ferroelectricity and magnetism from three-phonon coupling},
  author={Radaelli, Paolo G},
  journal={Physical Review B},
  volume={97},
  number={8},
  pages={085145},
  year={2018},
  publisher={APS}
}

@article{subedi2014,
  title={Theory of nonlinear phononics for coherent light control of solids},
  author={Subedi, Alaska and Cavalleri, Andrea and Georges, Antoine},
  journal={Physical Review B},
  volume={89},
  number={22},
  pages={220301},
  year={2014},
  publisher={APS}
}

@article{kampfrath2013,
  title={Resonant and nonresonant control over matter and light by intense terahertz transients},
  author={Kampfrath, Tobias and Tanaka, Koichiro and Nelson, Keith A},
  journal={Nature Photonics},
  volume={7},
  number={9},
  pages={680--690},
  year={2013},
  publisher={Nature Publishing Group UK London}
}

@article{shen1965,
  title={Theory of stimulated Brillouin and Raman scattering},
  author={Shen, Yo R and Bloembergen, N},
  journal={Physical Review},
  volume={137},
  number={6A},
  pages={A1787},
  year={1965},
  publisher={APS}
}

@article{rossi2021,
  title={Progress and challenges in ab initio simulations of quantum nuclei in weakly bonded systems},
  author={Rossi, Mariana},
  journal={The Journal of Chemical Physics},
  volume={154},
  number={17},
  year={2021},
  publisher={AIP Publishing}
}

@article{de2021,
  title={Colloquium: Nonthermal pathways to ultrafast control in quantum materials},
  author={De La Torre, Alberto and Kennes, Dante M and Claassen, Martin and Gerber, Simon and McIver, James W and Sentef, Michael A},
  journal={Reviews of Modern Physics},
  volume={93},
  number={4},
  pages={041002},
  year={2021},
  publisher={APS}
}

@article{muller1979,
  title={{$\mathrm{SrTiO_3}$}: An intrinsic quantum paraelectric below 4 K},
  author={M{\"u}ller, K Alex and Burkard, H},
  journal={Physical Review B},
  volume={19},
  number={7},
  pages={3593},
  year={1979},
  publisher={APS}
}

@article{han2007dielectric,
  title={Dielectric response of soft mode in ferroelectric {$\mathrm{SrTiO_3}$}},
  author={Han, Jiaguang and Wan, Fan and Zhu, Zhiyuan and Zhang, Weili},
  journal={Applied Physics Letters},
  volume={90},
  number={3},
  year={2007},
  publisher={AIP Publishing}
}

@article{ichikawa2005,
  title={Direct observation of the soft-mode dispersion in the incipient ferroelectric K Ta O 3},
  author={Ichikawa, Yuki and Nagai, Masaya and Tanaka, Koichiro},
  journal={Physical Review B—Condensed Matter and Materials Physics},
  volume={71},
  number={9},
  pages={092106},
  year={2005},
  publisher={APS}
}

@article{shapiro1972,
  title={Critical neutron scattering in {$\mathrm{SrTiO_3}$} and {$\mathrm{KMnF_3}$}},
  author={Shapiro, SM and Axe, JD and Shirane, G and Riste, T},
  journal={Physical Review B},
  volume={6},
  number={11},
  pages={4332},
  year={1972},
  publisher={APS}
}

@article{shirane1993,
  title={q dependence of the central peak in the inelastic-neutron-scattering spectrum of {$\mathrm{SrTiO_3}$}},
  author={Shirane, G and Cowley, RA and Matsuda, M and Shapiro, SM},
  journal={Physical Review B},
  volume={48},
  number={21},
  pages={15595},
  year={1993},
  publisher={APS}
}

@article{alicki1979,
  title={The quantum open system as a model of the heat engine},
  author={Alicki, Robert},
  journal={Journal of Physics A: Mathematical and General},
  volume={12},
  number={5},
  pages={L103},
  year={1979},
  publisher={IOP Publishing}
}

@article{talkner2020,
  title={Colloquium: Statistical mechanics and thermodynamics at strong coupling: Quantum and classical},
  author={Talkner, Peter and H{\"a}nggi, Peter},
  journal={Reviews of Modern Physics},
  volume={92},
  number={4},
  pages={041002},
  year={2020},
  publisher={APS}
}

@article{gonze1997,
  title={Dynamical matrices, Born effective charges, dielectric permittivity tensors, and interatomic force constants from density-functional perturbation theory},
  author={Gonze, Xavier and Lee, Changyol},
  journal={Physical Review B},
  volume={55},
  number={16},
  pages={10355},
  year={1997},
  publisher={APS}
}

@article{barra2021,
  title={Microcavity phonon polaritons from the weak to the ultrastrong phonon--photon coupling regime},
  author={Barra-Burillo, Mar{\'\i}a and Muniain, Unai and Catalano, Sara and Autore, Marta and Casanova, F{\`e}lix and Hueso, Luis E and Aizpurua, Javier and Esteban, Ruben and Hillenbrand, Rainer},
  journal={Nature communications},
  volume={12},
  number={1},
  pages={6206},
  year={2021},
  publisher={Nature Publishing Group UK London}
}

@article{barrett1952,
  title={Dielectric constant in perovskite type crystals},
  author={Barrett, John H},
  journal={Physical Review},
  volume={86},
  number={1},
  pages={118},
  year={1952},
  publisher={APS}
}

@article{shin2021,
  title={Quantum paraelectric phase of SrTiO 3 from first principles},
  author={Shin, Dongbin and Latini, Simone and Sch{\"a}fer, Christian and Sato, Shunsuke A and De Giovannini, Umberto and H{\"u}bener, Hannes and Rubio, Angel},
  journal={Physical Review B},
  volume={104},
  number={6},
  pages={L060103},
  year={2021},
  publisher={APS}
}

@article{hartmann2025,
  title={Intrinsic non-Markovian magnetisation dynamics},
  author={Hartmann, Felix and Unikandanunni, Vivek and Bargheer, Matias and Fullerton, Eric E and Bonetti, Stefano and Anders, Janet},
  journal={arXiv preprint arXiv:2512.07378},
  year={2025}
}

@article{anders2022,
  title={Quantum Brownian motion for magnets},
  author={Anders, Janet and Sait, CRJ and Horsley, Simon AR},
  journal={New Journal of Physics},
  volume={24},
  number={3},
  pages={033020},
  year={2022},
  publisher={IOP Publishing}
}

@article{strasberg2019,
  title={Non-Markovianity and negative entropy production rates},
  author={Strasberg, Philipp and Esposito, Massimiliano},
  journal={Physical Review E},
  volume={99},
  number={1},
  pages={012120},
  year={2019},
  publisher={APS}
}

@unpublished{Hartmann2025_inprep,
  author = {Felix Hartmann and Finja Tietjen and R. Matthias Geilhufe and Janet Anders},
  title  = {Entropy Production as Markovianity Measure in Magnetization Dynamics},
  note   = {Manuscript in preparation},
  year   = {2025}
}

\clearpage
\onecolumngrid
\appendix

\section{Quantum heat in driven open quantum systems}\label{appendix:heat}

We briefly summarize here a general framework for defining quantum heat in driven open quantum systems and derive the expression for the heat current used in the main text.

The total Hamiltonian is written as
\begin{align}
    H(t) = H_{\mathrm{ph}}(t) + H_B + H_{SB},
\end{align}
where $H_{\mathrm{ph}}(t)$ is the (possibly time-dependent) Hamiltonian of the driven system, $H_B$ is the Hamiltonian of the bath, and $H_{SB}$ describes the system–bath interaction. Throughout this Appendix we keep the notation $H_{\mathrm{ph}}(t)$ for the system Hamiltonian in order to be consistent with the main text, but the derivation is completely general.

The first law of thermodynamics for the system is expressed as
\begin{align}
    \langle\dot{U}_S\rangle = \langle\dot{W}\rangle + \langle\dot{\mathcal{Q}}\rangle,
\end{align}
where $\langle U_S\rangle$ is the internal energy of the system, $\langle W\rangle$ is the work performed by the external driving, and $\langle\mathcal{Q}\rangle$ denotes the heat flowing into the system. The internal energy is defined as
\begin{align}
    \langle U_S(t)\rangle
    = \mathrm{Tr}_S\!\left[H_{\mathrm{ph}}(t)\,\rho_S(t)\right]
    = \mathrm{Tr}_{SB}\!\left[H_{\mathrm{ph}}(t)\,\rho_{SB}(t)\right],
\end{align}
where $\rho_{SB}(t)$ is the density matrix of the full system (system plus bath) and
$\rho_S(t) = \mathrm{Tr}_B[\rho_{SB}(t)]$ is the reduced density matrix of the system. Taking the time derivative yields
\begin{align}\label{eq:dU}
    \langle\dot{U}_S\rangle
    &= \mathrm{Tr}_{SB}\!\left[\frac{\partial H_{\mathrm{ph}}(t)}{\partial t}\rho_{SB}(t)\right]
     + \mathrm{Tr}_{SB}\!\left[H_{\mathrm{ph}}(t)\frac{\partial\rho_{SB}(t)}{\partial t}\right].
\end{align}
By comparison with the first law, we identify the instantaneous power and heat flow as \cite{alicki1979}
\begin{align}
    \langle\dot{W}\rangle
    &= \mathrm{Tr}_{SB}\!\left[\frac{\partial H_{\mathrm{ph}}(t)}{\partial t}\rho_{SB}(t)\right],\\[1mm]
    \langle\dot{\mathcal{Q}}\rangle
    &= \mathrm{Tr}_{SB}\!\left[H_{\mathrm{ph}}(t)\frac{\partial\rho_{SB}(t)}{\partial t}\right].
\end{align}

The total energy of system plus bath is
\begin{align}
    \langle E_{\text{tot}}(t)\rangle
    &= \mathrm{Tr}_{SB}\!\left[H(t)\rho_{SB}(t)\right] \nonumber\\
    &= \mathrm{Tr}_{SB}\!\left[H_{\mathrm{ph}}(t)\rho_{SB}(t)\right]
     + \mathrm{Tr}_{SB}\!\left[H_B\rho_{SB}(t)\right]
     + \mathrm{Tr}_{SB}\!\left[H_{SB}\rho_{SB}(t)\right] \nonumber\\
    &= \langle U_S(t)\rangle + \langle E_B(t)\rangle + \langle E_{SB}(t)\rangle,
\end{align}
where $\langle E_B\rangle$ and $\langle E_{SB}\rangle$ denote the bath and interaction energies, respectively. Taking the time derivative and using the von Neumann equation $\dot{\rho}_{SB}(t) = -\tfrac{i}{\hbar}[H(t),\rho_{SB}(t)]$ one finds
\begin{align}\label{eq:dE1}
    \langle \dot{E}_{\text{tot}}\rangle
    &= \mathrm{Tr}_{SB}\!\left[\frac{\partial H(t)}{\partial t}\rho_{SB}(t)\right]
     + \mathrm{Tr}_{SB}\!\left[H(t)\frac{\partial\rho_{SB}(t)}{\partial t}\right] \nonumber\\
    &= \mathrm{Tr}_{SB}\!\left[\frac{\partial H_{\mathrm{ph}}(t)}{\partial t}\rho_{SB}(t)\right]
     = \langle\dot{W}\rangle,
\end{align}
since the second trace vanishes by cyclicity and only $H_{\mathrm{ph}}(t)$ is explicitly time dependent.

On the other hand, differentiating the decomposition of $\langle E_{\text{tot}}\rangle$ yields
\begin{align}\label{eq:dE2}
    \langle \dot{E}_{\text{tot}}\rangle
    &= \langle\dot{U}_S\rangle
     + \frac{d}{dt}\langle E_B\rangle
     + \frac{d}{dt}\langle E_{SB}\rangle \nonumber\\
    &= \langle\dot{W}\rangle + \langle\dot{\mathcal{Q}}\rangle
     + \frac{d}{dt}\langle E_B\rangle
     + \frac{d}{dt}\langle E_{SB}\rangle,
\end{align}
where in the second line we used Eq.~\eqref{eq:dU}. Introducing
\begin{align}
    \langle \tilde{\mathcal{Q}}\rangle
    = \langle E_B(t)\rangle - \langle E_B(t_0)\rangle
    = \mathrm{Tr}_{SB}\!\left[H_B\rho_{SB}(t)\right]
      - \mathrm{Tr}_{SB}\!\left[H_B\rho_{SB}(t_0)\right],
\end{align}
as the energy dissipated into the bath, we obtain from Eqs.~\eqref{eq:dE1} and \eqref{eq:dE2}
\begin{align}
    \langle\dot{\mathcal{Q}}\rangle
    = -\frac{d}{dt}\langle \tilde{\mathcal{Q}}\rangle
      - \frac{d}{dt}\langle E_{SB}\rangle.
\end{align}
Thus, the heat flowing into the system has two contributions: the change in the bath energy and the change in the interaction energy. In the weak-coupling limit,
\begin{align}
    H_{SB} \ll H_{\mathrm{ph}},\,H_B,
\end{align}
the interaction energy is much smaller than the system and bath energies, and its contribution to the energy balance can often be neglected. In this regime one approximately has
\begin{align}
    \langle\mathcal{Q}\rangle \approx -\langle \tilde{\mathcal{Q}}\rangle,
\end{align}
i.e., the heat entering the bath is approximately given by the negative of the heat entering the system.

For our purposes it is convenient to express the heat in terms of system operators only. From the first law,
\begin{align}\label{Eq:current}
    \langle\mathcal{Q}\rangle
    &= \Delta U_S - \langle W\rangle \nonumber\\
    &= \left[\langle U_S(\tau)\rangle - \langle U_S(t_0)\rangle\right]
       - \int_{t_0}^{\tau}\!dt\,\langle\dot{W}(t)\rangle \nonumber\\
    &= \mathrm{Tr}_{SB}\!\left[H_{\mathrm{ph}}(\tau)\rho_{SB}(\tau)\right]
       - \mathrm{Tr}_{SB}\!\left[H_{\mathrm{ph}}(t_0)\rho_{SB}(t_0)\right] \nonumber\\
    &\quad - \int_{t_0}^{\tau}\!dt\,\mathrm{Tr}_{SB}\!\left[\frac{\partial H_{\mathrm{ph}}(t)}{\partial t}\rho_{SB}(t)\right].
\end{align}
To connect this expression to the one used in the main text, we switch to the Heisenberg picture. Let $U(t)$ be the unitary time-evolution operator of the full system with $U(t_0)=\hat{I}$ and $\rho_{SB}(t) = U(t)\rho_{SB}(t_0)U^\dagger(t)$. Then
\begin{align}
    \tilde{H}_{\mathrm{ph}}(t)
    &= U^\dagger(t)\,H_{\mathrm{ph}}(t)\,U(t),\\[1mm]
    \widetilde{\frac{\partial H_{\mathrm{ph}}(t)}{\partial t}}
    &= U^\dagger(t)\,\frac{\partial H_{\mathrm{ph}}(t)}{\partial t}\,U(t),
\end{align}
are the corresponding Heisenberg-picture operators. Writing all expectation values with respect to the initial density matrix $\rho_{SB}(t_0)$, Eq.~\eqref{Eq:current} becomes
\begin{align}
    \langle\mathcal{Q}\rangle
    &= \int_{t_0}^{\tau}\!dt\,
       \left[
         \frac{d}{dt}\big\langle\tilde{H}_{\mathrm{ph}}(t)\big\rangle
         - \Big\langle
            \widetilde{\frac{\partial H_{\mathrm{ph}}(t)}{\partial t}}
           \Big\rangle
       \right] \nonumber\\
    &= \int_{t_0}^{\tau}\!dt\, J(t),
\end{align}
where the quantum heat current is defined as
\begin{align}\label{eq:Jt}
    J(t)
    = \frac{d}{dt}\big\langle\tilde{H}_{\mathrm{ph}}(t)\big\rangle
      - \Big\langle
         \widetilde{\frac{\partial H_{\mathrm{ph}}(t)}{\partial t}}
        \Big\rangle.
\end{align}
Here, $\langle\cdots\rangle$ denotes the ensemble average with respect to the full density matrix $\rho_{SB}(t_0)$. Equation~\eqref{eq:Jt} is the general expression for the heat current used in the main text; when applied to the specific phonon Hamiltonian of Eq.~\eqref{eq:Hph}, it leads to Eq.~\eqref{eq:J_t}.

\nocite{*}

\end{document}